\documentclass[3p]{elsarticle}

\usepackage{graphics}

\usepackage{amssymb}

\usepackage{comment}

\def\Journal#1#2#3#4{{#1} {\bf #2} (#4) #3}

\def\APJ{Astrophys. J.}

\def\JHEP{JHEP}

\def\JETPUSSR{JETP (USSR)}
 
\def\MPLA{Mod. Phys. Lett. A}

\def\NPB{Nucl. Phys. B}

\def\NPBSUPPL{Nucl. Phys. B. Proc. Suppl.}
\def\PLB{{Phys. Lett.} B}

\def\PLBOLD{Phys. Lett.}

\def\PRL{Phys. Rev. Lett.}
\def\PRD{Phys. Rev. D}

\def\PTP{Prog. Theor. Phys.}
\def\RMP{Rev. Mod. Phys.}

\def\SCIENCE{Science}

\journal{Physics Letters B}

\begin{document}

\begin{frontmatter}



\title{CP violation in bipair neutrino mixing}


\author{Teruyuki Kitabayashi\corref{cor1}}
\ead{teruyuki@keyaki.cc.u-tokai.ac.jp}

\author{Masaki Yasu\`{e}}
\ead{yasue@keyaki.cc.u-tokai.ac.jp}

\cortext[cor1]{Corresponding author}

\address{Department of Physics, Tokai University, 4-1-1 Kitakaname, Hiratsuka, Kanagawa, 259-1292, Japan}

\begin{abstract}
There are experimentally determined two best-fit points for the atmospheric neutrino mixing angle $\theta_{23}$:  $\sin^2 \theta_{23} = 0.413$ (case A) and 
$\sin^2 \theta_{23} = 0.594$ (case B). In the bipair neutrino mixing scheme, we predict $\sin^2\theta_{23} = \sqrt 2 -1$ (case 1) to be consistent with the case A and $\sin^2\theta_{23} = 2-\sqrt 2$ (case 2) to be consistent with the case B. If the case B is realized in nature, the bipair neutrino mixing provides a unique neutrino model consistent with the observation $\sin^2 \theta_{23} = 0.594$. However, the reactor neutrino mixing angle $\theta_{13}$ is predicted to be $\sin^2\theta_{13} = 0$, which is inconsistent with the observation. We propose a new modification scheme to yield $\sin^2\theta_{13} \neq 0$ utilizing the charged lepton contribution and study its effect on both of CP-violating Dirac and Majorana phases, which is numerically estimated. It is found that there appear striking differences between the case 1 and the case 2 in their phase structure.
\end{abstract}


\end{frontmatter}


\section{Introduction}
The results from the neutrino oscillation experiments have provided us with robust evidence that neutrinos have tiny masses and their flavor states are mixed with each other \cite{atmospheric, solar, reactor, accelerator}. A global analysis shows that the best-fit values of the mixing angles with $1\sigma$ allowed range are obtained as \cite{Gonzalez-Garcia2012}
\begin{eqnarray}
\sin^2 \theta_{12} &=& 0.302^{+0.013}_{-0.012} ,\nonumber \\
\sin^2 \theta_{23} &=& 0.413^{+0.037}_{-0.025} \ \oplus \ 0.594^{+0.021}_{-0.022},      \nonumber \\
\sin^2 \theta_{13} &=& 0.0227^{+0.0023}_{-0.0024},
\label{Eq:global}
\end{eqnarray}
where $\theta_{12}, \theta_{23}$ and $\theta_{13}$ stand for the solar, atmospheric and reactor neutrino mixing angle, respectively.\footnote{Another global analysis  is also reported \cite{FLMMPR2012, FTV2012}} There are two best-fit points for the atmospheric neutrino mixing angle $\theta_{23}$. We call these two best-fit points case A and case B as follows:
\begin{eqnarray}
\sin^2 \theta_{23} &=& 0.413 \quad {\rm (case \ A)}, \nonumber \\
\sin^2 \theta_{23} &=& 0.594 \quad {\rm (case \ B)}.
\label{Eq:observedAtmosphericAngle}
\end{eqnarray}

Neutrino mixings are described by the Pontecorvo-Maki-Nakagawa-Sakata mixing matrix $U_{PMNS}$ \cite{UMNS,PDG}, which is parameterized by three mixing angles $\theta_{12,23,13}$ as well as one CP-violating Dirac phase $\delta_{CP}$ and two CP-violating Majorana phases $\alpha_{2,3}$ \cite{CPphases} to be $U_{PMNS}=UK$ with
\begin{eqnarray}
&&
U = 
\left( {\begin{array}{*{20}{c}}
{{c_{12}}{c_{13}}}&{{s_{12}}{c_{13}}}&{{s_{13}}{e^{ - i{\delta _{CP}}}}}\\
{ -  {{s_{12}}{c_{23}} - {c_{12}}{s_{23}}{s_{13}}{e^{i{\delta _{CP}}}}} }&{{c_{12}}{c_{23}} - {s_{12}}{s_{23}}{s_{13}}{e^{i{\delta _{CP}}}}}&{{s_{23}}{c_{13}}}\\
{{s_{12}}{s_{23}} - {c_{12}}{c_{23}}{s_{13}}{e^{i{\delta _{CP}}}}}&{ - {{c_{12}}{s_{23}} - {s_{12}}{c_{23}}{s_{13}}{e^{i{\delta _{CP}}}}} }&{{c_{23}}{c_{13}}}
\end{array}} \right),
\nonumber\\
&&
K = {\rm diag.}(1, e^{i\alpha_2}, e^{i\alpha_3}),
\label{UPMNS}
\end{eqnarray}
where $c_{ij}=\cos\theta_{ij}$ and $s_{ij}=\sin\theta_{ij}$ $(i,j=1,2,3)$ \cite{PDG}. We will also use the notation of $t_{ij}=\tan\theta_{ij}$. 

There is a theoretical prediction of these mixing angles based on the bipair neutrino mixing scheme \cite{KitabayashiYasue2011}. In the bipair neutrino mixing scheme, there are two sets of solutions referred to as case 1 and case 2. We obtain $\sin^2\theta_{23}= \sqrt 2 -1 (=0.414)$ in the case 1, which is close to $\sin^2\theta_{23}$ in the case A, and $\sin^2\theta_{23} = 2 -\sqrt 2 (=0.586)$ in the case 2, which is close to $\sin^2\theta_{23}$ in the case B. At the same time, $\sin^2\theta_{12} = 1 - 1/\sqrt{2}(= 0.293)$ and $\sin^2\theta_{13} = 0$ are predicted. Although various other possibilities realizing the observed mixing angles $\theta_{12,23}$ have been proposed \cite{reviewObMixings, tribimaximal, transposedTribimaximal, bimaximal, goldenRatio1, goldenRatio2, hexagonal}, the bipair neutrino mixing can be a good candidate of neutrino mixing schemes for the both of the case A and the case B. Especially, if the case B is chosen as a right atmospheric neutrino mixing angle, the bipair neutrino mixing scheme is the only theoretical model that has a consistent prediction with the case B.

The bipair neutrino mixing predicts $\sin^2\theta_{13} = 0$ such as in Refs.\cite{tribimaximal, transposedTribimaximal, bimaximal, goldenRatio1, goldenRatio2, hexagonal}, which is inconsistent with the observation. It is expected that additional contributions to the mixing angles are produced by the charged lepton contributions \cite{chargedLeptonContributions} if some of the nondiagonal matrix elements of charged lepton mass matrix are nonzero so that the reactor mixing angle can be shifted to lie in the allowed region. Two modification schemes of the bipair neutrino mixing to induce $\sin^2\theta_{13} \neq 0$ to be consistent with the observation are discussed in Ref.\cite{KitabayashiYasue2012} and Ref.\cite{Kitabayashi2013}. The first modification scheme is not suitable to study effects of CP-violating Majorana phases while the second modification scheme is not suitable to study those of CP-violating Dirac phase. 

In this letter, we discuss third modification scheme based on the parameterization by Pascoli, Petcov and Rodejohann \cite{Pascoli2003} and evaluate charged lepton contribution to both of CP-violating Dirac and Majorana phases in the bipair neutrino mixing scheme.  We estimate sizes of these CP-violating phases, whose correlations are numerically obtained and discuss expected effects of CP violation in the case 1 and the case 2.

\section{Modified bipair neutrino mixing}
The bipair neutrino mixing is described by a mixing matrix $U$ with $\theta_{13}=0$, which is equipped with two pairs of identical magnitudes of matrix elements to be denoted by $U^0_{ij}$ ($i.j$=1,2,3). There are two cases of the bipair texture:

\begin{itemize}
\item 
The case 1 of the bipair neutrino mixing ($\vert U^0_{12} \vert = \vert U^0_{32} \vert$ and $\vert U^0_{22} \vert = \vert U^0_{23} \vert$) is parameterized by $U^0_{BP1}$ containing only one mixing angle $\theta_{12}$: 
\begin{eqnarray}
U^0_{BP1}=\left(
  \begin{array}{ccc}
    c      & s   & 0 \\
    -t^2  & t   & t \\
    st     & -s  & t/c 
  \end{array}
\right), 
\label{Eq:UnuBP1}
\end{eqnarray}
where $c=c_{12}$, $s=s_{12}$ and $t=t_{12}$. The mixing angles are predicted to be 
\begin{eqnarray}
\sin^2\theta_{12}^{BP1} &=& 1 - 1/\sqrt{2} = 0.293, \nonumber \\
\sin^2\theta_{23}^{BP1} &=& \sqrt{2} - 1 = 0.414,
\end{eqnarray}
and $\sin^2\theta_{13}^{BP1} = 0$. The case 1 of the bipair neutrino mixing well describes the 1$\sigma$ data of the solar neutrino mixing in Eq.(\ref{Eq:global}) and the case A of the atmospheric neutrino mixing in Eq.(\ref{Eq:observedAtmosphericAngle}).
\item 
The case 2 of the bipair neutrino mixing ($\vert U^0_{12} \vert = \vert U^0_{22} \vert$ and $\vert U^0_{32} \vert = \vert U^0_{33} \vert$) is parameterized by  $U^0_{BP2}$:
\begin{eqnarray}
U^0_{BP2}=\left(
  \begin{array}{ccc}
     c      & s   & 0 \\
     -st   & s   & t/c  \\
      t^2  & -t  & t 
  \end{array}
\right).
\label{Eq:UnuBP2}
\end{eqnarray}
The mixing angles are predicted to be
\begin{eqnarray}
\sin^2\theta_{12}^{BP2} &=& 1 - 1/\sqrt{2} = 0.293, \nonumber \\
\sin^2\theta_{23}^{BP2} &=& 2 - \sqrt{2} = 0.586,
\end{eqnarray}
and $\sin^2\theta_{13}^{BP2} = 0$. The case 2 of the bipair neutrino mixing well describes the 1$\sigma$ data of the solar neutrino mixing in Eq.(\ref{Eq:global}) and the case B of the atmospheric neutrino mixing in Eq.(\ref{Eq:observedAtmosphericAngle}).
\end{itemize}

Our modified bipair neutrino mixing is described by a specific type of $U_{PMNS}$, which will be constructed by the parameterization proposed by Pascoli, Petcov and Rodejohann \cite{Pascoli2003}. Namely, $U_{PMNS}$ is given by
\begin{eqnarray}
U_{PMNS} = \tilde{U}^0 K, \quad \tilde{U}^0 \equiv U_\ell^\dagger P U_\nu,
\end{eqnarray}
where $U_{\ell}$ (as well as $U_R$ to be used later) and $U_\nu$, respectively, arise from the diagonalization of the charged lepton mass matrix $M_\ell$ and of the neutrino mass matrix $M_\nu$ and $P$ is defined by
\begin{eqnarray}
P = {\rm diag.}(1, e^{i\phi_2}, e^{i\phi_3}).
\end{eqnarray}
The lepton mass matrices satisfy the relations of $M_\ell = U_\ell M_\ell^{diag} U_R^\dagger$ and $M_\nu = U_\nu M_\nu^{diag} U_\nu^T$ where $M_\ell^{diag}$ and $M_\nu^{diag}$ are, respectively, the diagonal mass matrix of the charged lepton and of the neutrino. 

The charged lepton mixing matrix $U_\ell$ can be parameterized by three mixing angles $\theta^\ell_{12,23,13}$ and one CP-violating Dirac phase $\delta$ as follows:
\begin{eqnarray}
U_\ell&=& 
\left( {\begin{array}{*{20}{c}}
{{c^\ell_{12}}{c^\ell_{13}}}&{{s^\ell_{12}}{c^\ell_{13}}}&{{s^\ell_{13}}{e^{ - i{\delta}}}}\\
{ -  {{s^\ell_{12}}{c^\ell_{23}} - {c^\ell_{12}}{s^\ell_{23}}{s^\ell_{13}}{e^{i{\delta}}}} }&{{c^\ell_{12}}{c^\ell_{23}} - {s^\ell_{12}}{s^\ell_{23}}{s^\ell_{13}}{e^{i{\delta}}}}&{{s^\ell_{23}}{c^\ell_{13}}}\\
{{s^\ell_{12}}{s^\ell_{23}} - {c^\ell_{12}}{c^\ell_{23}}{s^\ell_{13}}{e^{i{\delta}}}}&{ - {{c^\ell_{12}}{s^\ell_{23}} - {s^\ell_{12}}{c^\ell_{23}}{s^\ell_{13}}{e^{i{\delta}}}} }&{{c^\ell_{23}}{c^\ell_{13}}}
\end{array}} \right),
\label{Eq:generalUell}
\end{eqnarray}
where $c^\ell_{ij} = \cos\theta^\ell_{ij}$ and $s^\ell_{ij} = \sin\theta^\ell_{ij}$ $(i,j = 1,2,3)$.  To be more specific, we adapt the Cabibbo-Kobayashi-Maskawa like parameterization of the matrix $U_\ell$ \cite{Pascoli2003,Guinti_Tanimoto2002}, which is represented by small parameters having magnitude of the order of Wolfenstein parameter $\lambda \sim 0.227$ or less. In the small angle approximation, the charged lepton mixing matrix can be approximated to be:
\begin{eqnarray}
U_\ell&=& 
\left( {\begin{array}{*{20}{c}}
{1 - \frac{{\epsilon _{12}^2 + \epsilon _{13}^2}}{2}}&{{\epsilon _{12}}}&{{e^{ - i\delta }}{\epsilon _{13}}}\\
{ - {\epsilon _{12}} - {\epsilon _{23}}{\epsilon _{13}}{e^{i\delta }}}&{1 - \frac{{\epsilon _{12}^2 + \epsilon _{23}^2}}{2}}&{{\epsilon _{23}}}\\
{{\epsilon _{12}}{\epsilon _{23}} - {e^{i\delta }}{\epsilon _{13}}}&{ - {\epsilon _{23}} - {\epsilon _{12}}{\epsilon _{13}}{e^{i\delta }}}&{1 - \frac{{\epsilon _{23}^2 + \epsilon _{13}^2}}{2}}
\end{array}} \right),
\label{Eq:generalUell-approximated}
\end{eqnarray}
where $\epsilon_{ij} \equiv s^\ell_{ij}$. Since Eq.(\ref{Eq:generalUell-approximated}) is non-unitary, the resulting neutrino mixing matrix is not unitary.  Current data do not allow a big unitarity violation effect \cite{nonunitary}. We demand that the charged lepton mixing matrix is unitary within 1\%, which is roughly $\lambda^3$.  Therefore, we have retained the terms of the second order in $\epsilon_{ij}$ in Eq.(\ref{Eq:generalUell-approximated}). Although numerical calculations are performed  with approximate unitarity kept, results of our estimations will be shown within the first order in $\epsilon_{ij}$ to avoid apparent complexity of derived expressions.  We mention that there are other types of approximation for charged lepton mixing matrix \cite{chargedLeptonContributions}. 

If $M_\ell$ is non-diagonal, $U_\ell$ is associated with $M_\ell$.  We construct such a mass matrix that yields Eq.(\ref{Eq:generalUell-approximated}).  One possible form of $U_\ell$ arises from the wishful thinking of the similarity between the quark and lepton sectors \cite{Guinti_Tanimoto2002}. Moreover, with SU(5) GUT relation, we can make simple mass matrix of charged leptons \cite{Altarelli2004}. In minimal SU(5) GUT, the charged lepton mass matrix $M_\ell$ and the down quark mass matrices $M_d$ are related each other to satisfy the relation $M_\ell = M^T_d$; therefore, we expect $U_R = U_d$. The Cabibbo-Kobayashi-Maskawa matrix $U_{CKM}$ for quark sector is given by $U_{CKM} = U_u^\dagger U_d$. Given that the quark mixing angles are small, either both $U_u$ and $U_d$ are nearly diagonal or they are nearly equal. One possible choice is that $U_d$ is nearly diagonal and $U_R$ is also nearly diagonal. In this case, the mass matrix $M_\ell$ for with $M_\ell^{diag} = {\rm diag.}(m_e, m_\mu, m_\tau)$ is approximately given by 
\begin{eqnarray}
M_\ell = U_\ell M_\ell^{diag} U_R^\dagger \sim \left(
  \begin{array}{ccc}
     m_e            & m_\mu \epsilon_{12}   & m_\tau e^{-i\delta}\epsilon_{13} \\
     -m_e \epsilon_{12}  & m_\mu  & m_\tau \epsilon_{23}  \\
      -m_e e^{i\delta}\epsilon_{13}       & -m_\mu \epsilon_{23}  & m_\tau 
  \end{array}	
\right),
\label{Eq:Mell}
\end{eqnarray}
where $U_R \sim 1$, where terms of the first order in $\epsilon_{ij}$ are retained for the sake of simplicity. This is one example of the charged lepton mass matrix, which is consistent with Eq.(\ref{Eq:generalUell-approximated}).

To estimate the charged lepton contributions to the mixing angles in the bipair neutrino mixing, we take the neutrino mixing matrix to be either $U_\nu = U_{BP1}^0$ or $U_\nu = U_{BP2}^0$ . As a definition of the bipair neutrino mixing, there is no phase parameter in $U_{BP1}^0$ and $U_{BP2}^0$. The mixing matrix $U$ has three real parameters $\epsilon_{12}, \epsilon_{13}, \epsilon_{23}$ and five phases $\delta, \phi_2, \phi_3, \alpha_2, \alpha_3$. 

In the case of $U_\nu$ = $U_{BP1}^0$, we obtain the mixing matrix $\tilde{U}^0_{BP1}$ for $\tilde{U}^0$ as 
\begin{eqnarray}
\tilde{U}^0_{BP1} &=&\left(
  \begin{array}{ccc}
     c    & s  & 0 \\
     -e^{i\phi_2}t^2 & e^{i\phi_2}t & e^{i\phi_2}t  \\
      e^{i\phi_3} st  & -e^{i\phi_3} s & e^{i\phi_3} t/c
  \end{array}
\right) 
\nonumber\\
&&+
\left(
  \begin{array}{ccc}
     e^{i\phi_2}t^2\epsilon_{12} - \tilde{s} t \epsilon_{13}    & -e^{i\phi_2}t \epsilon_{12} + \tilde{s} \epsilon_{13}  & -e^{i\phi_2}t \epsilon_{12} -  \frac{\tilde{s}}{c^2}\epsilon_{13} \\
     c\epsilon_{12} -e^{i\phi_3} st \epsilon_{23} & s\epsilon_{12} + e^{i\phi_3} s\epsilon_{23} & -e^{i\phi_3} \frac{t}{c} \epsilon_{23} \\
     e^{i\delta} c\epsilon_{13} - e^{i\phi_2}t^2 \epsilon_{23} & e^{i\delta} s\epsilon_{13} + e^{i\phi_2}t \epsilon_{23} & e^{i\phi_2}t \epsilon_{23}
  \end{array}
\right), 
\label{Eq:UInBP1}
\end{eqnarray}
where we define $\tilde{s} = e^{-i(\delta - \phi_3)}s$. From $U_{PMNS}=\tilde{U}^0_{BP1} K$ and the following general relations
\begin{eqnarray}
\sin^2\theta_{12} = \frac{\vert U_{12} \vert^2}{\vert U_{11} \vert^2 + \vert U_{12} \vert^2}, \quad
\sin^2\theta_{23} = \frac{\vert U_{23} \vert^2}{\vert U_{23} \vert^2 + \vert U_{33} \vert^2}, \quad
\sin^2\theta_{13} = \vert U_{13} \vert^2,
\label{Eq:leptonMixingAngles}
\end{eqnarray}
where $U_{ij}$=$(U_{PMNS})_{ij}$ ($i.j$=1,2,3), the mixing angles are obtained as follows:
\begin{eqnarray}
\sin^2\theta_{12} &=& \left( 1-2^{5/4} \epsilon_{12}\cos\phi_2+ 2\epsilon_{13} \cos(\delta - \phi_3) \right) \sin^2\theta_{12}^{BP1}, \nonumber \\
\sin^2\theta_{23} &=& \left( 1-2^{5/4} \epsilon_{23} \cos(\phi_2-\phi_3) \right) \sin^2\theta_{23}^{BP1}, \nonumber \\
\sin^2\theta_{13} &=& \left( \epsilon_{12}^2 + 2^{5/4}\epsilon_{12}\epsilon_{13} \cos(\delta + \phi_2 -\phi_3) \right) \sin^2\theta_{23}^{BP1} 
+ 2 \epsilon_{13}^2 \sin^2\theta_{12}^{BP1}. 
\label{Eq:leptonMixingAnglesInBP1}
\end{eqnarray}

In the case of $U_\nu$ = $U^0_{BP2}$, we obtain the mixing matrix $\tilde{U}^0_{BP2}$ as 
\begin{eqnarray}
\tilde{U}^0_{BP2} &=&\left(
  \begin{array}{ccc}
     c    & s  & 0 \\
     -e^{i\phi_2}st & e^{i\phi_2}s & e^{i\phi_2}t/c  \\
      e^{i\phi_3} t^2  & -e^{i\phi_3} t & e^{i\phi_3} t
  \end{array}
\right) 
\nonumber\\
&&+
\left(
  \begin{array}{ccc}
     e^{i\phi_2}st\epsilon_{12} - \tilde{t} t \epsilon_{13}    & -e^{i\phi_2}s \epsilon_{12} + \tilde{t} \epsilon_{13}  & -e^{i\phi_2}\frac{t}{c} \epsilon_{12} -  \tilde{t}\epsilon_{13} \\
     c\epsilon_{12} -e^{i\phi_3} t^2 \epsilon_{23} & s\epsilon_{12} + e^{i\phi_3} t\epsilon_{23} & -e^{i\phi_3} t \epsilon_{23} \\
     e^{i\delta} c\epsilon_{13} - e^{i\phi_2}st \epsilon_{23} & e^{i\delta} s\epsilon_{13} + e^{i\phi_2}s \epsilon_{23} & e^{i\phi_2}\frac{t}{c} \epsilon_{23}
  \end{array}
\right), 
\label{Eq:UInBP2}
\end{eqnarray}
where we define $\tilde{t} = e^{-i(\delta - \phi_3)}t$. The mixing angles from $U_{PMNS}=\tilde{U}^0_{BP2} K$ and Eq.(\ref{Eq:leptonMixingAngles}) are obtained as follows:
\begin{eqnarray}
\sin^2\theta_{12} &=& \left( 1- 2 \epsilon_{12} \cos\phi_2 + 2^{5/4} \epsilon_{13} \cos(\delta - \phi_3) \right) \sin^2\theta_{12}^{BP2}, \nonumber \\
\sin^2\theta_{23} &=& \left( 1 - 2^{3/4} \epsilon_{23} \cos(\phi_2-\phi_3) \right) \sin^2\theta_{23}^{BP2}, \nonumber \\
\sin^2\theta_{13} &=& \sqrt{2}\left( \epsilon_{13}^2 + 2^{5/4}\epsilon_{12}\epsilon_{13} \cos(\delta + \phi_2 -\phi_3) \right) \sin^2\theta_{12}^{BP2} 
+ \epsilon_{12}^2 \sin^2\theta_{23}^{BP2}. 
\label{Eq:leptonMixingAnglesInBP2}
\end{eqnarray}
%

\section{CP-violating phases}
There are rephasing invariant quantities for CP-violating Dirac phase as well as CP-violating Majorana phases.  It is known that CP-violating Dirac phase $\delta_{CP}$ is determined by the Jarlskog invariant $J$ \cite{Jarlskog1985} to be:
\begin{eqnarray}
\sin\delta_{CP} = \frac{{\rm Im} \left(U_{11}U_{22}U_{12}^\ast U_{21}^\ast \right)}{c_{12}c_{23}c_{13}^2s_{12}s_{23}s_{13}}.
\label{Eq:sinDeltaCP}
\end{eqnarray}
One can also find that CP-violating Majorana phases $\alpha_{2,3}$ are determined to be:
\begin{eqnarray}
\alpha_2= {\rm arg} \left(U_{12}U_{11}^\ast\right), \quad  \alpha_3= {\rm arg} \left(U_{13}U_{11}^\ast\right) + \delta_{CP},
\label{Eq:alpha2AndAlpha3}
\end{eqnarray}
which can be used to predict CP-violating Majorana phases from $U_{PMNS}$.

To find Eq.(\ref{Eq:alpha2AndAlpha3}), e.g., rephasing invariant expressions for Majorana phases $\alpha_{2,3}$, it is useful to employ general parameterization of $U$ instead of Eq.(\ref{UPMNS}), which is containing three Dirac phases $\delta$, $\rho$ and $\tau$ as well as one additional phase $\gamma$ \cite{Pascoli2003,generalU}:
\begin{eqnarray}
U &=& 
\left( {\begin{array}{*{20}{c}}
1&0&0\\
0&{{e^{i\gamma }}}&0\\
0&0&{{e^{ - i\gamma }}}
\end{array}} \right)\left( {\begin{array}{*{20}{c}}
1&0&0\\
0&{{c_{23}}}&{{s_{23}}{e^{i\tau }}}\\
0&{ - {s_{23}}{e^{ - i\tau }}}&{{c_{23}}}
\end{array}} \right)
\left( {\begin{array}{*{20}{c}}
{{c_{13}}}&0&{{s_{13}}{e^{ - i\delta }}}\\
0&1&0\\
{ - {s_{13}}{e^{i\delta }}}&0&{{c_{13}}}
\end{array}} \right)
\nonumber\\
&&\cdot\left( {\begin{array}{*{20}{c}}
{{c_{12}}}&{{s_{12}}{e^{i\rho }}}&0\\
{ - {s_{12}}{e^{ - i\rho }}}&{{c_{12}}}&0\\
0&0&1
\end{array}} \right),
\label{U-Ours}
\end{eqnarray}
and three Majorana phases $\varphi_{1,2,3}$, 
\begin{eqnarray}
Q =
  \left(
  \begin{array}{ccc}
     e^{i\varphi_1}   & 0                    & 0 \\
     0                     & e^{i\varphi_2} & 0 \\
     0                     & 0                   & e^{i\varphi_3} 
  \end{array}
  \right),
\label{Q-Ours}
\end{eqnarray}
giving $U_{PMNS} = UQ$.  After ${\delta _{CP}}$ is taken to be ${\delta _{CP}} = \delta  + \rho  + \tau $, we readily reach
\begin{eqnarray}
&&
U_{PMNS} = P^\prime
\left( {\begin{array}{*{20}{c}}
{{c_{12}}{c_{13}}}&{{s_{12}}{c_{13}}}&{{s_{13}}{e^{ - i{\delta _{CP}}}}}\\
{ - \left( {{c_{23}}{s_{12}} + {s_{23}}{c_{12}}{s_{13}}{e^{i{\delta _{CP}}}}} \right)}&{{c_{23}}{c_{12}} - {s_{23}}{s_{12}}{s_{13}}{e^{i{\delta _{CP}}}}}&{{s_{23}}{c_{13}}}\\
{{s_{23}}{s_{12}} - {c_{23}}{c_{12}}{s_{13}}{e^{i{\delta _{CP}}}}}&{ - \left( {{s_{23}}{c_{12}} + {c_{23}}{s_{12}}{s_{13}}{e^{i{\delta _{CP}}}}} \right)}&{{c_{23}}{c_{13}}}
\end{array}} \right)Q^\prime,
\label{UPMNS-Ours}
\end{eqnarray}
where
\begin{eqnarray}
P^\prime &=&
\left( {\begin{array}{*{20}{c}}
{{e^{i\rho }}}&0&0\\
0&{{e^{i\gamma }}}&0\\
0&0&{{e^{ - i\left( {\gamma  + \tau } \right)}}}
\end{array}} \right),
\quad
Q^\prime =
\left( {\begin{array}{*{20}{c}}
{{e^{i\beta_1 }}}&0&0\\
0&{{e^{i\beta_2 }}}&0\\
0&0&{e^{ i\beta_3}}
\end{array}} \right),
\label{Pprime}
\end{eqnarray}
with ${\beta _1} = {\varphi _1} - \rho$, ${\beta _2} = {\varphi _2}$ and ${\beta _3} = {\varphi _3} + \tau$.

CP-violating Dirac phase in the form of the rephasing invariant quantity, we can make the following product as one possible choice:
\begin{eqnarray}
U_{11}U_{22}U_{12}^\ast U_{21}^\ast,
\label{DeltaInv}
\end{eqnarray}
which is independent of the redundant phases as well as $\alpha_{2,3}$. The Jarlskog invariant $J$ is expressed to be: 
\begin{eqnarray}
J =  {\rm Im} \left( {{U_{11}}{U_{22}}U_{12}^\ast U_{21}^\ast  } \right),
\label{Jarlskog}
\end{eqnarray}
giving
\begin{eqnarray}
{\rm Im} \left( {{U_{11}}{U_{22}}U_{12}^\ast  U_{21}^\ast  } \right) = {c_{12}}{c_{23}}c_{13}^2{s_{12}}{s_{23}}{s_{13}}\sin {\delta _{CP}}.
\label{J}
\end{eqnarray}
In the similar way, it is possible to describe CP-violating Majorana phases in terms of rephasing invariant quantities.  In fact, using Eq.(\ref{UPMNS-Ours}), one can find that 
${U_{12}}U_{11}^\ast$, ${U_{13}}U_{11}^\ast$ and ${U_{13}}U_{12}^\ast$ \cite{RephasingInvariance}, which are independent of redundant phases, lead to
\begin{eqnarray}
{U_{12}}U_{11}^\ast  &=& {s_{12}}{c_{13}}{c_{12}}{c_{13}}{e^{i\left( {{\beta _2} - {\beta _1}} \right)}},
\nonumber\\
{U_{13}}U_{11}^\ast  &=& {s_{13}}{c_{12}}{c_{13}}{e^{i\left( {{\beta _3} - {\beta _1} - {\delta _{CP}}} \right)}},
\label{InvarianMajoranaPhase}\\
{U_{13}}U_{12}^\ast  &=& {s_{13}}{s_{12}}{c_{13}}{e^{i\left( {{\beta _3} - {\beta _2} - {\delta _{CP}}} \right)}},
\nonumber
\end{eqnarray}
from which
\begin{eqnarray}
{\beta _2} - {\beta _1} &=& \alpha_2 = \arg \left( {{U_{12}}U_{11}^\ast } \right),
\nonumber\\
{\beta _3} - {\beta _1} &=&  \alpha_3 = \arg \left( {{U_{13}}U_{11}^\ast } \right) + {\delta _{CP}},
\label{InvarianPhi}\\
{\beta _3} - {\beta _2} &=& \arg \left( {{U_{13}}U_{12}^\ast } \right) + {\delta _{CP}},
\nonumber
\end{eqnarray}
can be derived.  

\section{Numerical analysis}
We have performed the random search to find the parameter set of $\{\epsilon_{12,13,23}, \delta, \phi_2, \phi_3, \delta_{CP},\alpha_2, \alpha_3\}$. To enhance the predictability, we assume that $\epsilon_{12} = \epsilon_{13} = \epsilon_{23} = \epsilon$. The unitarity of $U_\ell$ dictates that
\begin{eqnarray}
U_\ell ^\dag {U_\ell } = I + \left( {\begin{array}{*{20}{c}}
0&0&{ - {e^{ - i\delta }}}\\
0&{2\cos \delta }&0\\
{ - {e^{i\delta }}}&0&0
\end{array}} \right){\epsilon ^3},
\label{Unitarity}
\end{eqnarray}
for $\epsilon^3 \le {\mathcal O}(0.01)$. The mixing matrices $\tilde{U}^0_{BP1,BP2}$ are accordingly estimated to be:
\begin{eqnarray}
{\tilde U}^0 &=& \left( {\begin{array}{*{20}{c}}
c&s&0\\
{ - {e^{i{\phi _2}}}s{c_{23}}}&{{e^{i{\phi _2}}}c{c_{23}}}&{{e^{i{\phi _2}}}{s_{23}}}\\
{{e^{i{\phi _3}}}s{s_{23}}}&{ - {e^{i{\phi _3}}}c{s_{23}}}&{{e^{i{\phi _3}}}{c_{23}}}
\end{array}} \right)
\nonumber\\
&+& \left( {\begin{array}{*{20}{c}}
{s\left( {{e^{i{\phi _2}}}{c_{23}} - {e^{ - i\left( {\delta  - {\phi _3}} \right)}}{s_{23}}} \right)}&{ - c\left( {{e^{i{\phi _2}}}{c_{23}} - {e^{ - i\left( {\delta  - {\phi _3}} \right)}}{s_{23}}} \right)}&{ - \left( {{e^{i{\phi _2}}}{s_{23}} + {e^{ - i\left( {\delta  - {\phi _3}} \right)}}{c_{23}}} \right)}\\
{c - {e^{i{\phi _3}}}s{s_{23}}}&{s + {e^{i{\phi _3}}}c{s_{23}}}&{ - {e^{i{\phi _3}}}{c_{23}}}\\
{{e^{i\delta }}c - {e^{i{\phi _2}}}s{c_{23}}}&{{e^{i\delta }}s + {e^{i{\phi _2}}}c{c_{23}}}&{{e^{i{\phi _2}}}{s_{23}}}
\end{array}} \right)\epsilon 
\nonumber\\
&+& \left( {\begin{array}{*{20}{c}}
{ - \left( \begin{array}{l}
c - {e^{ - i\left( {\delta  - {\phi _2}} \right)}}s{c_{23}}\\
 - {e^{i{\phi _3}}}s{s_{23}}
\end{array} \right)}&{ - \left( \begin{array}{l}
s + {e^{ - i\left( {\delta  - {\phi _2}} \right)}}c{c_{23}}\\
 + {e^{i{\phi _3}}}c{s_{23}}
\end{array} \right)}&{{e^{ - i\left( {\delta  - {\phi _2}} \right)}}{s_{23}} - {e^{i{\phi _3}}}{c_{23}}}\\
{s\left( {{e^{i{\phi _2}}}{c_{23}} - {e^{ - i\left( {\delta  - {\phi _3}} \right)}}{s_{23}}} \right)}&{ - c\left( {{e^{i{\phi _2}}}{c_{23}} - {e^{ - i\left( {\delta  - {\phi _3}} \right)}}{s_{23}}} \right)}&{ - \left( {{e^{i{\phi _2}}}{s_{23}} + {e^{ - i\left( {\delta  - {\phi _3}} \right)}}{c_{23}}} \right)}\\
{ - {e^{i{\phi _3}}}s{s_{23}}}&{{e^{i{\phi _3}}}c{s_{23}}}&{ - {e^{i{\phi _3}}}{c_{23}}}
\end{array}} \right){\epsilon ^2},
\nonumber\\
\label{U-ell-nu}
\end{eqnarray}
where ${c_{23}} = t/s,{s_{23}} = t$ for $\tilde{U}^0_{BP1}$ and ${c_{23}} = t,{s_{23}} = t/c$ for $\tilde{U}^0_{BP2}$. The input parameters are $\epsilon, \delta, \phi_2, \phi_3$ while the output parameters are CP-violating phases $\delta_{CP}, \alpha_2, \alpha_3$. We vary the input parameters in the range of $0 \le \epsilon \le 0.227$ and $-\pi \le \delta, \phi_2, \phi_3 \le \pi$. 
\begin{figure}[t]
 \includegraphics{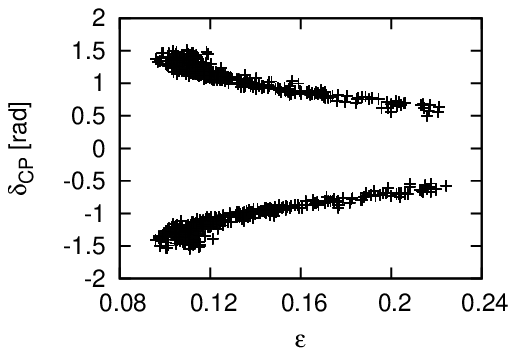}
 \includegraphics{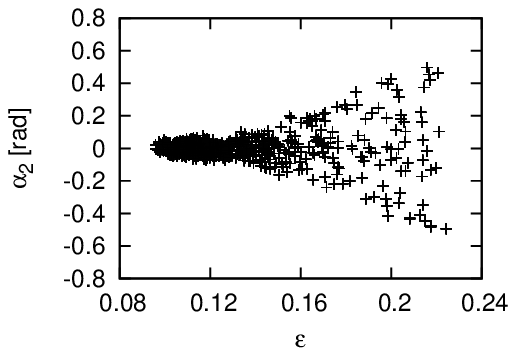}
 \includegraphics{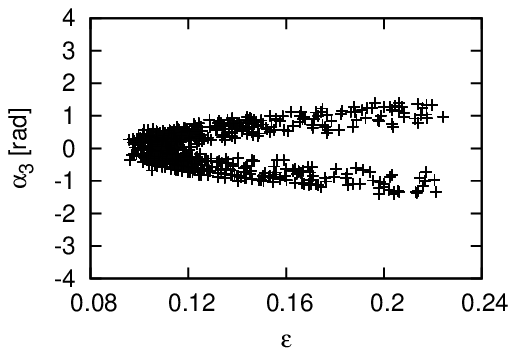}
 \caption{\label{fig:fig1} The dependence of $\delta_{CP}$ and $\alpha_{2,3}$ on $\epsilon$ in the case 1.}
 \includegraphics{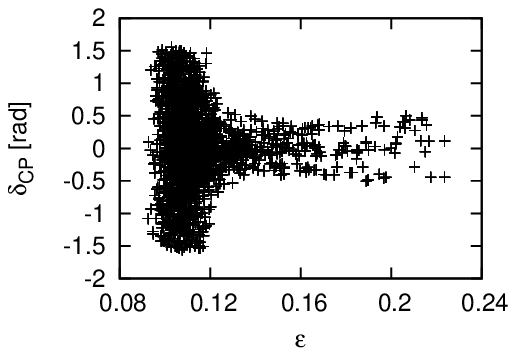}
 \includegraphics{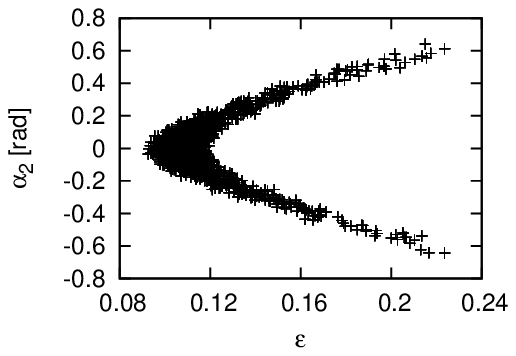}
 \includegraphics{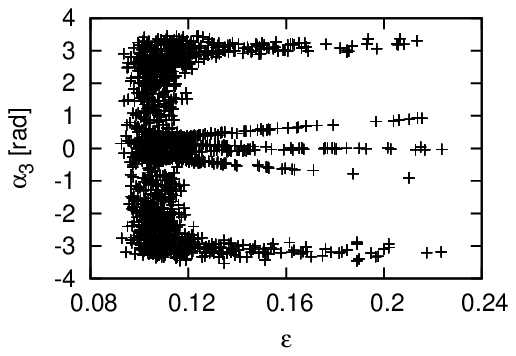}
 \caption{\label{fig:fig2} Same as Fig.\ref{fig:fig1} but in the case 2.}
\end{figure}
\begin{figure}[t]
 \includegraphics{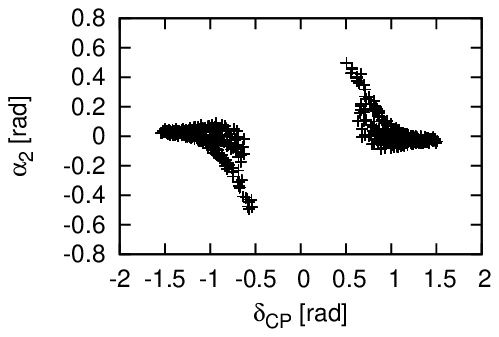}
 \includegraphics{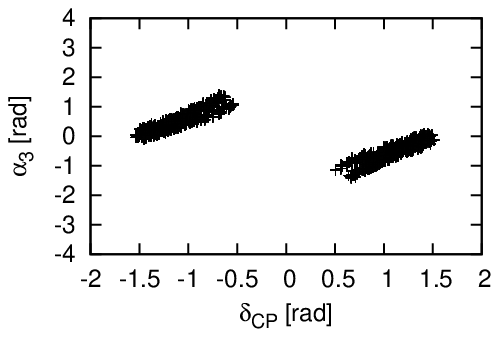}
 \includegraphics{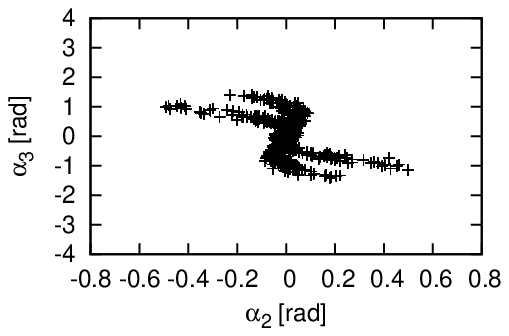}
 \caption{\label{fig:fig3} The mutual dependence of $\delta_{CP}$ and $\alpha_{2,3}$ in the case 1.}
 \includegraphics{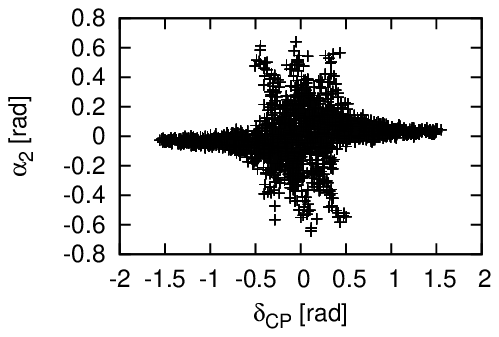}
 \includegraphics{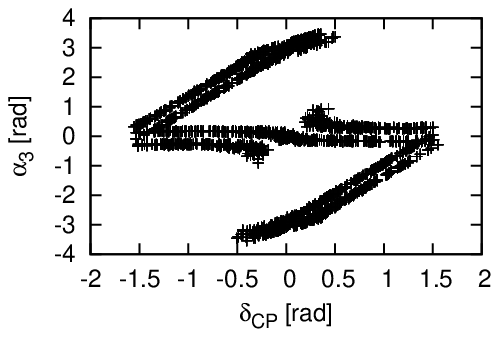}
 \includegraphics{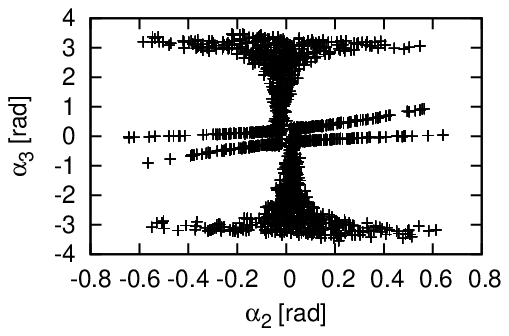}
 \caption{\label{fig:fig4} Same as Fig.\ref{fig:fig3} but in the case 2.}
\end{figure}

For the case 1, we have searched the consistent parameter set with the 1$\sigma$ data of the solar and reactor neutrino mixing angles in Eq.(\ref{Eq:global}) and of the case A of the atmospheric mixing angle in Eq.(\ref{Eq:observedAtmosphericAngle}), e.g., $\sin^2 \theta_{12} = 0.302^{+0.013}_{-0.012}$, $\sin^2 \theta_{23} = 0.413^{+0.037}_{-0.025}$, and $\sin^2 \theta_{13} = 0.0227^{+0.0023}_{-0.0024}$. Fig.\ref{fig:fig1} shows the result of our numerical analysis in the case 1 of the modified bipair neutrino mixing: 
\begin{enumerate}
\item The $\delta_{CP} - \epsilon$ plot shows that\footnote{The range of $\delta_{CP}$ is taken to be $-\pi/2\sim\pi/2$.}
\begin{itemize}
\item $\delta_{CP}$ is constrained to be $\vert \delta_{CP}\vert \gtrsim 0.5$;
\item $\delta_{CP}$ is constrained to be $\vert\delta_{CP}\vert\gtrsim 1.0$ for the minimal value of $\epsilon$ around 0.1;
\item the smaller magnitude of $\delta_{CP}$ is allowed for the larger value of $\epsilon$.
\end{itemize}
\item The $\alpha_2 - \epsilon$ plot shows that 
\begin{itemize}
\item $\alpha_2$ is constrained to be $\vert \alpha_2\vert \lesssim 0.5$;
\item $\alpha_2$ is constrained to be $\vert\alpha_2\vert\lesssim 0.1$ for the smaller value of $\epsilon$ satisfying $0.09 \lesssim \epsilon \lesssim 0.14$;
\item the larger magnitude of $\alpha_2$ is allowed for the larger value of $\epsilon$.
\end{itemize}
\item The $\alpha_3 - \epsilon$ plot shows that 
\begin{itemize}
\item $\alpha_3$ is constrained to be $\vert \alpha_3\vert \lesssim \pi/2$;
\item $\alpha_3$ is constrained to be $\vert\alpha_3\vert\lesssim 0.5$ for the minimal value of $\epsilon$ around 0.1;
\item the larger magnitude of $\alpha_3$ is allowed for the larger value of $\epsilon$. 
\end{itemize}
\end{enumerate}

For the case 2, we have searched the consistent parameter set  with the 1$\sigma$ data of the solar and reactor neutrino mixing angles in Eq.(\ref{Eq:global}) and of the case B of the atmospheric mixing angle in Eq.(\ref{Eq:observedAtmosphericAngle}), e.g., $\sin^2 \theta_{12} = 0.302^{+0.013}_{-0.012}$, $\sin^2 \theta_{23} = 0.594^{+0.021}_{-0.022}$, and $\sin^2 \theta_{13} = 0.0227^{+0.0023}_{-0.0024}$. Fig.\ref{fig:fig2} shows the result of our numerical analysis in the case 2 of the modified bipair neutrino mixing:
\begin{enumerate}
\item The $\delta_{CP} - \epsilon$ plot shows that 
\begin{itemize}
\item $\delta_{CP}$ is constrained to be $\vert \delta_{CP}\vert \lesssim 0.5$ for $\epsilon \gtrsim 0.12$;
\item no constraint on $\delta_{CP}$ arises for $\epsilon \lesssim 0.12$;
\item the smaller magnitude of $\delta_{CP}$ is allowed in the broad range of $0.09 \lesssim \epsilon \lesssim 0.22$.
\end{itemize}
\item The $\alpha_2 - \epsilon$ plot shows that 
\begin{itemize}
\item $\alpha_2$ is constrained to be $\vert \alpha_2\vert \lesssim 0.6$;
\item the smaller magnitude of $\alpha_2$ is allowed for $0.09 \lesssim \epsilon \lesssim 0.12$;
\item the larger magnitude of $\alpha_2$ is allowed for the larger value of $\epsilon$.
\end{itemize}
\item The $\alpha_3 - \epsilon$ plot shows that 
\begin{itemize}
\item $\alpha_3$ is constrained to be $\vert \alpha_3\vert \lesssim \pi$;
\item $\alpha_3$ is constrained to be either $\alpha_3\sim 0$ around its minimal value or $\vert \alpha_3\vert \sim \pi$ around its maximal value in the broad range of $0.12 \lesssim \epsilon \lesssim 0.22$;
\item $\alpha_3$ takes any allowed values for $0.09 \lesssim \epsilon \lesssim 0.12$.
\end{itemize}
\end{enumerate}

It is also useful to present correlations among measurable quantities for the direct comparison with future experimental data of CP-violating phases. Shown in Fig.\ref{fig:fig3} and Fig.\ref{fig:fig4} are mutual dependence of CP-violating phases $\delta_{CP}$ and $\alpha_{2,3}$. 

For the case 1, we see the following mutual dependence in Fig.\ref{fig:fig3}:
\begin{enumerate}
\item The $\alpha_2 - \delta_{CP} $ plot shows that $\vert \alpha_2\vert \lesssim 0.5$ and that
\begin{itemize}
\item $\vert \alpha_2 \vert$ approaches to its maximal value around 0.5 as $\vert \delta_{CP} \vert$ approaches its minimal value around 0.5;
\item $\vert \alpha_2 \vert$ approaches to its minimal value around 0 as $\vert \delta_{CP} \vert$ approaches its maximal value around $\pi/2$;
\end{itemize}
\item The $\alpha_3 - \delta_{CP}$ plot shows that $\vert \alpha_3\vert \lesssim \pi/2$ and that
\begin{itemize}
\item $\vert \alpha_3\vert$ approaches to its maximal value of $\pi/2$ as $\vert \delta_{CP}\vert$ approaches to its minimal value around 0.5;
\item $\vert \alpha_3\vert$ approaches to its minimal value of 0 as $\vert \delta_{CP}\vert$ approaches to its maximal value of $\pi/2$;
\item roughly speaking, $\alpha_3$ is scattered around the straight line of $\alpha_3 = 3(2\delta_{CP}\pm\pi)/4$.
\end{itemize}
\item The $\alpha_3 - \alpha_2$ plot shows that 
\begin{itemize}
\item $\vert \alpha_2\vert$ approaches to its maximal value around 0.5 for $\vert\alpha_3\vert\sim 1$;
\item $\alpha_3$ is constrained to be $\vert \alpha_3\vert\lesssim 1.0$ for $\vert \alpha_2\vert\lesssim 0.1$.
\end{itemize}
\end{enumerate}
On the other hand, for the case 2, we see the following mutual dependence in Fig.\ref{fig:fig4}:
\begin{enumerate}
\item The $\alpha_2 - \delta_{CP} $ plot shows that $\vert \alpha_2\vert \lesssim 0.6$ and 
\begin{itemize}
\item $\delta_{CP}$ takes any allowed values for $\alpha_2 \sim 0$;
\item $\alpha_2$ takes any allowed values for $\vert \delta_{CP} \vert \lesssim 0.5$.
\end{itemize}
\item The $\alpha_3 - \delta_{CP}$ plot shows that
\begin{itemize}
\item $\vert \alpha_3\vert$ approaches to its maximal value of $\pi$ as $\vert \delta_{CP}\vert$ approaches to its minimal value of 0;
\item $\vert \alpha_3\vert$ approaches to its minimal value of 0 as $\vert \delta_{CP}\vert$ approaches to its maximal value of $\pi/2$;
\item roughly speaking, $\alpha_3$ is scattered around the straight line of $\alpha_3 = 2\delta_{CP}\pm\pi$ as well as around $\alpha_3=0$;
\item $\delta_{CP}$ takes any allowed values for $\alpha_3\sim 0$,
\end{itemize}
\item The $\alpha_3 - \alpha_2$ plot shows that 
\begin{itemize}
\item $\alpha_3$ is constrained to be either $\alpha_3\sim 0$ around its minimal value or $\vert \alpha_3\vert \sim \pi$ around its maximal value for $0.2 \lesssim \vert \alpha_2 \vert \lesssim 0.6$;
\item $\alpha_2$ ($\alpha_3$) takes any allowed values for $\alpha_3 \sim 0$ ($\alpha_2 \sim 0$).
\end{itemize}
\end{enumerate}
It can be stated that the case 1 is excluded for $\vert \delta_{CP}\vert \lesssim 0.3$ and that the case 2 is excluded for $\vert \delta_{CP}\vert \gtrsim 0.5$ if $\vert \alpha_2\vert \gtrsim 0.2$.

\section{Summary and discussions}
%
The bipair neutrino mixing has predicted the solar neutrino mixing angle $\theta_{12}$  and the atmospheric neutrino mixing angle $\theta_{23}$ to be consistent with the currently observed experimental ones: 
\begin{itemize}
\item 
For $\theta_{12}$, our prediction is $\sin^2\theta_{12} = 0.293$  to be consistent with the best-fit value of $\sin^2 \theta_{12} = 0.302$.
\item 
For $\theta_{23}$, our predictions are $\sin^2\theta_{23} = 0.414$ in the case 1  to be consistent with the best-fit value of $\sin^2 \theta_{23} = 0.413$ (case A) and $\sin^2\theta_{23} = 0.586$  in the case 2  to be consistent with the best-fit value of $\sin^2 \theta_{23} = 0.594$ (case B).  
\end{itemize}
For the reactor neutrino mixing angle $\theta_{13}$, our prediction of $\sin^2\theta_{13} = 0$ is inconsistent with the best-fit value of $\sin^2 \theta_{13} = 0.0227$. We have discussed effects on neutrino mixings from the charged lepton contribution, which induce appropriate sizes of $\theta_{13}$ as well as CP-violating Dirac phase $\delta_{CP}$ as a measurable quantity. CP-violating Majorana phases $\alpha_2$ and $\alpha_3$ are determined by $U_{PMNS}$ as rephasing invariant quantities to be
$\alpha_2 = \arg \left( {{U_{12}}U_{11}^\ast } \right)$ and $\alpha_3 = \arg \left( {{U_{13}}U_{11}^\ast } \right) + {\delta _{CP}}$.

>From numerical calculations based on the simplification of $\epsilon_{12} = \epsilon_{13} = \epsilon_{23} = \epsilon$, where $\epsilon_{12,13,23}$ characterizes three types of charged lepton contributions, we have found that there are striking differences between the case 1 and the case 2 in the relation of $\delta_{CP}$ - $\epsilon$, $\alpha_2$ - $\epsilon$ and $\alpha_3$ - $\epsilon$ in Fig.\ref{fig:fig1} and Fig.\ref{fig:fig2} as well as in the relation of $\alpha_2 - \delta_{CP}$, $\alpha_3 - \delta_{CP}$ and $\alpha_3 - \alpha_2$ in Fig.\ref{fig:fig3} and Fig.\ref{fig:fig4}. We can use these differences to predict CP-violating phases for comparison with the results of the future global analysis. For example, if the case A of the global analysis for the atmospheric neutrino mixing angle $\sin^2 \theta_{23} = 0.413$ to be a correct mixing angle, we can expect that the case 1 is realized and that the following predictions are obtained: 
\begin{enumerate}
\item 
The larger CP-violating Dirac phase around $\vert \delta_{CP}\vert \sim \pi/2$ favors for
\begin{itemize}
\item 
the suppressed charged lepton contribution of $\epsilon \lesssim 0.1$ as shown in Fig.\ref{fig:fig1};
\item
the suppressed CP-violating Majorana phase $\alpha_3$ around $\alpha_3 \sim 0$ as shown in Fig.\ref{fig:fig3}.
\end{itemize}
\item 
The smaller CP-violating Dirac phase around $\vert \delta_{CP}\vert \sim 0.5$ (as the maximally allowed value) favors for
\begin{itemize}
\item 
the larger charged lepton contribution of $\epsilon \sim 0.23$ as shown in Fig.\ref{fig:fig1};
\item
the larger CP-violating Majorana phase $\alpha_3$ around $\vert\alpha_3\vert \sim \pi/2$ (as the maximally allowed value) as shown in Fig.\ref{fig:fig3}.
\end{itemize}
\item $\alpha_3$ is scattered around the straight line of $\alpha_3 = 3(2\delta_{CP}\pm\pi)/4$.
\end{enumerate}
On the other hand, if the case B of the global analysis for the atmospheric neutrino mixing angle $\sin^2 \theta_{23} = 0.594$ to be a correct mixing angle, we can expect that the case 2 is realized and that the following predictions are obtained: 
\begin{enumerate}
\item The larger CP-violating Dirac phase around $\vert \delta_{CP}\vert \sim \pi/2$ favors is possible for 
\begin{itemize}
\item 
the suppressed charged lepton contribution of $\epsilon \lesssim 0.12$ as shown in Fig.\ref{fig:fig2};
\item
the suppressed CP-violating Majorana phase $\alpha_3$ around $\alpha_3 \sim 0$ as shown in Fig.\ref{fig:fig4}.
\end{itemize}
\item The smaller CP-violating Dirac phase $\vert \delta_{CP}\vert \lesssim 0.5$ favors for 
\begin{itemize}
\item 
the mild charged lepton contribution of $\epsilon \gtrsim 0.12$ as shown in Fig.\ref{fig:fig2};
\item
the suppressed CP-violating Majorana phase $\alpha_3$ around $\alpha_3 \sim 0,\pm\pi$ as shown in Fig.\ref{fig:fig4}.
\end{itemize}
\item $\alpha_3$ is scattered around the straight line of $\alpha_3 = 2\delta_{CP}\pm\pi$ as well as staying around $\alpha_3 \sim 0$.
\end{enumerate}
We can also observe that the case 1 is excluded for $\vert \delta_{CP}\vert \lesssim 0.5$ while the case 2 is excluded for $\vert \delta_{CP}\vert \gtrsim 0.5$ if $\vert \alpha_2\vert \gtrsim 0.2$.





\bibliographystyle{elsarticle-num}
\bibliography{<your-bib-database>}



\end{document}